**Title**

'napari-toska: Topological Skeleton Analysis for Network-Based Shape Representation and Object Evaluation'

**Authors, Affiliations & Contact(s)**

Allyson Quinn Ryan[1,2,3,†], Johannes Soltwedel[3], and Carl D. Modes[1,2,3,*,**]

[1] Max Planck Institute for Molecular Cell Biology and Genetics, Pfotenhauerstraße 108, 01307 Dresden, Saxony, Germany

[2] Center for Systems Biology Dresden, Pfotenhauerstraße 108b, 01307 Dresden, Saxony, Germany

[3]Cluster of Excellence Physics of Life, Arnoldstraße 18, 01307 Dresden, Saxony, Germany

[*] Lead contact

[**] Corresponding Author

[†] Current Affiliation: Chan Zuckerberg Biohub San Francisco, 499 Illinois Street, San Francisco, California 94158, United States of America

[*,**] Correspondence: modes@mpi-cbg.de

**Summary**

Shape analysis and classification are popular methods for biologists, biophysicists and mathematicians investigating relationships between object function and form. Classic shape descriptors, such as sphericity, can be powerful but may be insufficient for more complex shapes. Here, we present 'napari-toska' a topological skeleton based method to analyze complex objects by representing their shape asymmetries as networks. Using global neighborhood principles, classic network science metrics and spatial feature embedding we create instance segmentation object profiles to be used for immediate or downstream classification. napari-toska can also follow temporal dynamics and identify network features capable of differentiating between experimental phenotypes. We incorporated the capacity to measure absolute spatial features of objects to bring in aspects of scale. Furthermore, napari-toska identifies certain segmentation errors through the emergence or loss of network cycles. Combined, napari-toska functions allow for flexible and in-depth shape profiling of intricate shapes often observed in biological and physical settings where robust, yet precise, system configuration is essential to downstream processes.

**Introduction**

How and why an object's form is critical to its function is a long standing and highly relevant question in biology. Understanding the principles of final shape generation and a shape's unique identifiers, allows researchers to make strides toward answering this and related questions. To this end, shape description and classification are common tools for mathematicians, physicists, biologists and engineers. Shape description methodologies vary greatly in their foundations and consequently may not be, or at least currently are not, easily applicable to a broad range of datatypes.

Before delving deeper into the fields of shape description, image processing and network science, we would like to inform readers that we have collated a glossary of terms that we believe are useful and/or critical to the comprehension of both the background and results presented (Appendix A, bold italicized text). Some terms are unique to this study while others are used frequently in their corresponding discipline. It should also be noted that while multiple definitions exist for several glossary terms, we have only included those relevant to this study in an effort to limit any potential misunderstandings.

Well established mathematical definitions and groupings (*e.g.,* contour or region-based) of shape descriptors exist (Zhang and Lu, 2004). For example, sphericity is a region-based descriptor that compares an object's mass distribution to that of a perfect sphere by taking a defined ratio of the object's volume and surface area (Zhang and Lu, 2004; Pincus and Theriot, 2007; Lian et al., 2013). ***Sphericity*** (see Glossary), ***roundness*** and many other shape descriptors assess an object's shape in reference to one or more regular shapes (Wadell, 1935; Bribiesca, 2000; Hryciw et al., 2016; Maroof et al., 2020). In 3D, such metrics require accurate measurement of an object's surface area and/or volume. The ease of use and accuracy of these shape descriptors depends heavily on if the object(s) of interest are represented in ***continuous space*** or ***discrete space*** (Figure 1A).

While volume assessment is a relatively simple and robust calculation in both continuous and discrete datasets, surface area approximation is prone to large errors in discrete space. Perimeter and surface area approximation methods for objects stored in discrete space formats include Convex Hull calculation and Marching Cube algorithms (Barber et al., 1996; Lorenson and Cline, 1987; Lewiner et al., 2003). Both methods render triangulated meshes , which are semi-continuous surfaces, but differ in the inclusion or exclusion of an object's concavities. Using such methods, fitting the formula for a sphere or ellipsoid is relatively simple for uniformly shaped objects with little population level variation (*e.g.*, oranges) is relatively simple. In these cases, metrics like sphericity are capable of detecting significant shape changes or subpopulation(s) (*e.g.*, sumo versus navel oranges).

Unfortunately, in disciplines like developmental and cell biology where shape data is commonly stored in discrete space (*i.e.*, microscopy images), shapes are often more convoluted and comparison to regular shapes is insufficient (Alves-Afonso et al., 2021). This is further complicated by the fact that when applying surface approximation methods can easily cause shape information to be lost (e.g., over-smoothing) or misrepresented without proper parameterization, which can be tedious and costly, both in terms of time and computational resources. Several robust 2D contour-based descriptors exist to describe irregular objects; however, in many cases the definitions either cannot or have not been extended to 3D (Iivarinen and Visa, 1996; Iivarinen et al., 1997). One such method is Curvature Scale Space, which converts an object's boundaries into a line profile that can be used for multi-object comparison (Mokhtarian et al., 1997; Mokhtarian and Suomela, 1998; Abbasi et al., 1999; and Lu, 2004; Frejlichowski, 2012). This can be considered to be a form of dimensionality reduction, which is useful for several reasons including accelerating processing time.

With this information in mind, we asked if and how we could circumvent the requirement of robust and accurate surface area approximation while also reducing the dimensionality of 3D objects to understand complex shapes. We chose to develop a ***topological skeleton*** analysis method for both 2D and 3D digitized datasets. From the field of ***digital topology***, topological skeletons are an extension of the ***Medial Axis Transform*** (MAT) for application to digital datasets. Classically defined in continuous space, the MAT 'burns away' parts of an object that are not essential to its core shape. This leaves behind a set of ***points*** and/or ***lines***, often resembling tree branching, that reflect an object's base shape asymmetry. This is achieved using the grassfire transform, which is effectively a type of ***distance transform*** (Blum, 1967; Blum 1973; Leymarie and Levine, 1992; Kats and Pizer, 2003). This proves useful for understanding seemingly irregular and unpredictable shapes (Blum 1967; Blum 1973). However, the emergence of digital topology brought to light a limitation of the MAT. When applied to discretized objects, MAT will over detect sharp angles 'corners' and therefore yield an overly branched result. The development of ***thinning algorithms*** resolved this problem. It should be noted that thinning algorithms can differ not only in computational speed and cost, but also in definitions of topological preservation (Lam et al., 1992; Sato et al., 2000; Serino et al., 2011). Topological skeletons are the union of the MAT and a thinning algorithm. For the concerns of this paper, we will consider the Lee method due to its robustness in preserving object topology and reflection of surface variability (Lee et al., 1994), though the methods described in this paper are generally applicable to other skeleton generating algorithms as well (*e.g.,* 2D Zhang skeletons) (Zhang and Suen, 1984).

In addition to retaining an object's shape asymmetry basis and preserving object topology, topological skeletons offer the significant advantage of reducing dataset spatial dimensionality to one in both 2D and 3D. This dramatically reduces downstream analysis computational time. Furthermore, as with Curvature Scale Space, dimensionality reduction increases the flexibility of object-object comparison. For this reason, as well as retaining information relevant to features such as surface concavities, topological skeletons have been applied to a wide range of studies including object recognition and quantification of pathological phenotypes (Serino et al., 2014; Nunez-Iglesias et al., 2018). In developmental biology, simple topological skeleton feature extraction is capable of differentiating between spatially adjacent cell populations where traditional molecular methods could not (Alves-Afonso et al., 2021).

Here, we cross topological skeleton generation with ***network science*** methods. Using a simple skeletal parsing algorithm based on complete neighborhood assessment (Figure 1B-D), we are able to extract unique object features and dynamics from image ***instance segmentations*** without the need for parameter fitting or surface rendering methods. This is a highly desirable feature as it circumvents the necessity for user-defined parameters during analysis phases and therefore the introduction of user bias. By subsequently processing parsed skeletons as ***undirected network graphs*** that can be weighted by absolute spatial measurements (Figure 1E), we bring together the fields of shape description and network science allowing for a more in-depth statistical evaluation of objects. In order for a broad range of researchers to apply our analyses to their digitized datasets, we implemented them as a plugin for napari (Sofroniew et al., 2022) called napari-toska in which flexible pipelines can be easily designed and executed by users in either the napari GUI or greater Python-based workflows.

## Results

Network conversion of topological skeletons simplifies downstream analyses

We established a pipeline to covert pixelated/voxelated topological skeletons into undirected graphs using a simple set of ***neighborhood*** relationships (Figure 1B-E). First, each ***pixel/voxel*** of a skeleton is evaluated using the appropriate ***kernel***. Some thinning and topological skeleton algorithms permit the use of the $N_4$ kernel to evaluate pixel neighborhoods in 2D (Figure 1B,C), which only considers pixels sharing an ***edge*** as possible neighbors and results in more rigid skeletons due to unnecessarily thick regions (*i.e.,* widths greater than a single pixel). In 3D, the parallel is the use of either the $N_6$ or $N_{18}$ kernels, which consider only ***face***-sharing or face/edge-sharing pixels as possible neighbors respectively. To avoid these more rigid skeletons, we have exclusively used the Lee algorithm to generate topological skeletons, which is based on what we refer to as complete neighborhood kernels (Figure 1B,C; Lee et al., 1994). In 2D, this is the $N_8$ kernel, which allows a pixels sharing either an edge or ***vertex*** with the pixel of interest to be considered a neighbor (Figure 1B,C). In 3D, the $N_{26}$ kernel defines possible neighboring pixels to be any that share a face, edge or vertex with the voxel of interest (Figure 1B,C).

Following evaluation, pixels/voxels are classified as either ***endpoints***, ***chain points*** or ***branch points*** based on their neighbor counts (*i.e.*, endpoints have a single neighbor, chain points have two and branch points have at least three; Figure 1D,E). ***Branches***, which are stretches of neighboring chain points, are bounded by two end or branch points in any combination (Figure 1E). To create the undirected graph, we treat the end and branch points as ***network nodes*** and the branches as ***network edges*** (Figure 1E). At this stage, the network nodes and network edges are unweighted, meaning no component of the network is 'favored' over another. In order to have the network more accurately reflect the skeleton, and by extension the object's shape and size, we allow network nodes and/or network edges to be weighted by absolute measures extracted from either the instance segmentation (*e.g.*, branch length) or by using the instance segmentation to index into the original image dataset (*e.g.*, pixel intensity). With a ***weighted network***, we could now ask questions such as how to traverse a skeleton to observe the most of the object's shape asymmetry. This is achieved by following what we have termed the ***spine***, which is the longest, shortest path between any two endpoints in the network without edge repetition (Figure 1E). If calculated on an unweighted network, several paths could potentially be the spine as it would simply be the number of branch point nodes or network edges used to move from one endpoint node to another, but by weighting the network the spine becomes a proxy for an axis of variability through the object. Depending on how the network is weighted, various aspects of an object's skeleton can be evaluated. Some of these are discussed in later sections.

<u>Polygons reveal emergent elongation and branching behaviors of topological skeletons</u>

We first used this pipeline to analyze topological skeletons of several reference shapes with increasing complexity to build a foundational understanding of skeleton behavior before moving to open access, curated 2D instance segmentations of multiple cell types (Maška et al., 2014; Ulman et al., 2017). We started with ***regular polygons***, which are simple, 2D, convex objects with simple equations for perimeter and are that allowed us to explore the effects of moving from continuous to discrete space representation. We found that regular polygons' topological skeletons fall into three categories – henceforth termed ***Class 1***, ***Class 2*** and ***Class 3*** (Figure 2A; Movies S1-3). These terms relate to the number of vertices in the regular polygon in continuous space. Class 1 topological skeletons, which are generated from regular polygons where the number of vertices is divisible by 4, reduce to the single pixel that approximates the polygon's centre of mass (Figure 2A left; Movie S1). It should be noted that this can be considered a degenerate state of topological skeletons. While these states do exist and to a certain extent can be unravelled by mapping additional information onto the skeleton, such techniques are outside the scope of this paper. Class 2 skeletons, generated from any regular polygon where the number of vertices is divisible by 2 but not 4, possess a single branch whose length is proportional to some function of the polygon's internal angles and side length (Figure 2A top right; Movie S2). Finally, Class 3 skeletons are generated from regular polygons with an odd number of vertices. Class 3 skeletons are the simplest 2D objects in which skeletal branching can be observed (Figure 2A bottom right; Movie S3). In order to understand how the discretization (*i.e.*, pixel) resolution impacts variability in topological skeleton generation, we examined skeleton size (*i.e.*, pixel count) relative to regular polygon side length over a range of circumscribing radii (*i.e.*, the minimal radius of a circle that can completely surround a polygon or object). As expected, this ratio is subject to greater variation for all three classes at smaller values of circumscribing radii. This implies that a dataset's pixel resolution is lower than necessary to correctly assess object shape or repetitively achieve the same analytical result under noisy conditions. We also observe that this ratio quickly approaches 0 for Class 1 skeletons, which follows logically given that these skeletons are comprised of a single pixel regardless of circumscribing radius (Figure 2B). Interestingly, we find that the ratio for Class 2 and 3 skeletons approach distinct values as circumscribing radius increases (Figure 2B). These values are  likely linked to the elongation and branching behaviors of each class.

An important limitation of topological skeleton generation emerges when considering the simplest Class 3 regular polygon, the equilateral triangle (Figure 2A). It's topological skeleton is indistinguishable from its continuous MAT counterpart. The pixels representing the equilateral triangle's continuous space vertices cannot be removed by a thinning algorithm due to the algorithms inability to interpret any acute angles less than 45° within a single pixel's $N_8$ neighborhood. In unrotated discrete representations of regular polygons, this limitation only exists for the equilateral triangle (Movie S1-3). However, if rotated in discrete space, the problem of true vertex detection extends to all three regular polygon classes and is exacerbated when the ratio of pixel resolution to polygon side length increases (Movie S4-6). If uni-axial elongation is introduced into a Class 1 regular polygon (*e.g.,* a square) and subsequently rotated in digital space to induce the vertex over-detection, the resultant topological skeletons exhibits the robust behavior of growing and shrinking branches in correlation with rotational angle (Figure 2C; Movie S7). The limitation of true or false vertex interpretation at intervals of 45° is not limited to objects with perimeters containing sharp, angular transitions in trajectory. Similar rotationally induced branching behaviors with a periodicity of 45° is also observed in increasingly complex, curved objects with curvature sign changes (Movie S8). This consistent behavior shows that topological skeleton branching emergence is limited by pixel/voxel resolution and the discrete nature of images, not object complexity.

<u>Significant changes in curvature induce skeletal branching in objects with concavities</u>

Continuously curved objects with multiple ***radii of curvature*** (*e.g.*, an ellipse) and objects with surface concavities offer greater challenges for topological skeleton analysis and shape classification. For the sake of focus, we will only discuss trends relating to skeletal elongation and the emergence of branching. Let us consider the skeletons of uni-axially stretched Class 1 regular polygons and ellipses (Figure 2D). Such objects are in essence an extension of Class 2 regular polygons as they have non-branching, elongated skeletons. While in Class 2 regular polygons skeleton length is related to the number of vertices, ellipse skeleton length reflects the ***minor-major axis length ratios*** (Figure 2D-E; Movie S9). There is a linear relationship between the minor-major axis length ratio and skelton-major axis length ratio (Figure 2E). The greater the difference in major axis length and minor axis length, the longer an ellipse's skeleton will be (Figure 2D-E; Movie S9).

Thus far, we have only considered convex shapes. How do topological skeletons behave when perimeter/surface concavities are introduced? To address this in 2D, we analyzed two shape groups — crescent moon-like shapes and Pac-Man-like shapes (Figure 2F,G). For simplicity we generate these shapes by overlapping perfect circles (see Methods for details). Interestingly, the skeletons of crescent moon-like shapes behave like a hybrid of Class 2 regular polygons and the equilateral triangle of Class 3 (Figure 2A,F). Like Class 2, a crescent moon-like shape's skeleton is a single, non-branching line; like the equilateral triangle, the skeleton reaches all the way to the object's perimeter (Figure 2F). These features are independent its shape generation ratio (see Movie S10, Methods and Glossary Appendix for details). Here, like with polygons, the skeleton's extent to the perimeter reflects an acute angle that is induced by the changing sign of curvature where the perimeter transitions from a convex to a concave region. These features, curved branches and corner extensions, are also present in Pac-Man-like skeletons' (Figure 2G, Movie S11). However, unlike in crescent moon-like skeletons, Pac-Man-like skeletons exhibit shape generation ratio dependent branching (Figure 2G-H, Movie S11, Methods and Glossary Appendix). These results imply that while shapes may contain large perimeter/surface concativites, if the curvature radii of the convex and concave regions are similar enough, the induced shape asymmetry is not great enough to warrant the emergence of topological skeletal branching. Similarly, if a perimeter concavity is sufficiently shallow and therefore does not induce significant shape asymmetry, neither skeletal branching nor elongation will not occur (Figure 2H; Movie S11).

<u>Population and object-specific dynamics are tractable through skeletal feature extraction</u>

With the limitations and robust behaviors of topological skeletons now characterized, we turned to biological data, which is rich in dynamic shape change, to examine more complex shapes and determine what shape features napari-toska can detect. Open access, curated 2D instance segmentations of motile, non-confluent *in vitro* cells were used to generate, parse, spine extract and network convert topological skeletons as described in previous sections (Figure 1A, 3A-D; Maška et al., 2014; Ulman et al., 2017). With these parsings and networks, users can evaluate both single cell and population level skeleton dynamics. Thanks to the open access spatio-temporal tracking data that accompanied the instance segmentations, we were able to follow an individual cell's skeleton dynamics over time (Figure 3E). Single cell dynamics can be compared to population level means and standard deviations to pinpoint a timepoint subset(s) that indicate statistically significant deviations from baseline cell behavior (Figure 3E-J). The population level metrics can also be used to simulate the average cell's topological skeleton network (Figure 3K). Potentially, these average networks could act as a computationally simple foundation over which shape change dynamics can be simulated and compared in *in silico* experiments between cell types or other experimental conditions.

<u>3D asymmetry characterization and non-spherical object topology detection</u>

Extending napari-toska to datasets with higher dimensions reinforces previously observed relationships of object symmetry. Simple convex, curved 3D objects (*i.e.*, spheres and ellipsoids) have simple, non-branching skeletons comprised of a single, straight line like their 2D counterparts (Figure 4A-A', Movie S12). Also like their 2D counterparts, there is a linear relationship between ellipsoid minor-major axis length ratio and skeletal elongation (Figure 4B). The central difference between ellipses and ellipsoids is that there can be two minor axis lengths (Figure 4B). In a different case of skeletal elongation in 3D, shapes such as those of budding yeast can be decomposed into overlapping spheres (Figure 4C; Movie S13). If a single, straight line can pass through the origin points of all overlapping spheres, the object's skeleton will be a line segment no longer than the line segment existing between the two furthest sphere origin points (Figure 4C'-D). The relative size difference and degree of overlapping between spheres in budding yeast is reflected in the length of their skeletons (Figure 4D). This again parallels 2D counterparts. We wondered if, in analogy to 2D shapes with multiple radii of curvature, introduction of surface concavities with significant curvature mismatch in 3D skeletal branching would be emerge in a similar manner.

We extended our analysis to 3D objects with non-spherical topologies (*e.g.,* volumes with completely internalized voids) and observed interesting skeletal behaviors. For the sake of simplicity, we generated what we henceforth refer to as 'semi-hollow spheres', which are perfect spherical volumes with an internal void that is also a perfect sphere (Figure 4E''-E'''; Movie S14). Depending on the position of the void, skeletons display varying branching complexity. If a void is positioned at the volume's edge such that the void's surface and the volume's surface intersect, the resultant skeleton is highly branched (Figure 4E'). If the void's center is not identical to the sphere's and the bounding surfaces do not intersect, the semi-hollow sphere's skeleton will be a surface rather than a collection of lines. This surface will contain multiple radii of curvature to reflect the positional difference between the sphere and the void (Figure 4E''). If the object and its void are identical in shape and share a centre of mass, the resultant skeleton will be of identical shape and with a size averaging the object and the void (Figure 4E'''). However, the skeleton will have no volume, that is to say, is hollow. The further from the object's centre of mass the void is shifted, or the curvature mismatch is lessened, the simpler the skeleton becomes (Figure 4E-F). As expected, the extensive branching that emerges when a void's bounding surface breaches a volume's bounding surface is limited to a specific range of shape generation ratios (Figure 4H, see Methods for shape generation ratio definitions). Interestingly, a power law like relationship between branch counts and normalized skeleton size is observable over at least one decade in surface intersecting semi-hollow spheres and spheres with significant surface concavities (Figure 4G). This may be indicative of a governing behavior of skeletons dictated by volume surface roughness (*i.e.*, discretization resolution) and radii or curvature. It should be noted that in the semi-hollow sphere cases shown here, because both volume and void are convex, the skeletal surface(s) will also be convex. If surfaces and voids were to contain significant surface concavities, it is possible for resultant skeletal surfaces to as well, but this is outside the scope of this study.

Comparison of 3D biological datasets

To understand more complex 3D objects, we again turned to biological data. An open access timelapse of a single cancer cell embedded in Matrigel (Maška et al., 2014; Ulman et al., 2017) provided an ideal opportunity to test if napari-toska is capable of differentiating between experimental conditions. The same study also created and published a simulated dataset. The simulated data was intended to replicate several aspects of fluorescence microscopy, such as pixel level noise and sub-cellular signal distribution. We used the experimental dataset (Figure 5A,A') as a basis of comparison to evaluate how accurately the simulated dataset (Figure 5B,B') was able to replicate cell shape dynamics. Before proceeding to downstream analyses, we confirmed that the simulation preserves the same topology as the experiment. Next, as with the 2D instance segmentation datasets, we used the edge-weighted network conversions of topological skeletons to extract spines (Figure 5A',B'). We observed that, while the experimental cell's spine length was prone to fluctuations around a steady mean over time, the simulated cell's spine length steadily increased over time with little variation between timepoints (Figure 5C).

Visually, a user may assume the source of this difference to be from cell surface projection morphology (Figure 5A,B). To test this hypothesis, we tracked the endpoint-branch point count ratio for both the experimental and simulated datasets over time (Figure 5C). Interestingly, while the ratio for the simulated cell's network is initially quite high, it quickly drops to a similar value to that of the experimental cell's network (Figure 5C). At first glance, these appear to be conflicting results as to whether or not the experiment and simulation can be distinguished from one another. However, when viewed in combination, the two skeletal analyses indicate that there is a hidden, critical and missing parameter from the simulation. This parameter should couple the shape complexity of  surface projections with their life-death rates. Because cell volume conservation also needs to be observed, this parameter could also potentially limit spine length range. As it is not within the scope of this paper, we did not explicitly solve for and test the validity of such a parameter, but these skeletal analysis results already offer directions for simulation improvement to better recapitulate cancer cell shape dynamics.

We next probed if and how sensitively napari-toska would detect shape differences between cell compartments. We chose to compare nuclear instance segmentations topological skeletal networks to those of membrane instance segmentations in 3D (Figure 5D-E'; Allen Institute for Cell Science, 2018). We found that, indeed, topological skeleton networks are significantly different. While there are many features that could be compared, the temporal differences in endpoint-branch count ratio are particularly interesting. Nuclear networks exhibit a ratio above or near equal to 1, but with a high degree of variability (Figure 5F). This is indicative of simple, but unstable network structure. In contrast, membrane network endpoint-branch count ratios show a consistent ratio of approximately 0.5 with low variability, which implies a relatively complex but highly stable network (Figure 5F). Futhermore, one could speculate that there exists a two state system due to clear reduction in nuclear network ratio variability in the back half of timelapse. Together, these trends could describe a system where small force fluctuations are experienced at the sub-cellular (i.e., nuclear) level but do not cause significant impact at the multi-cellular level. Further intracellular network comparison(s) could be highly useful in confluent or packed cell systems used to study rigidity/fluidity.

<u>Background skeletal features correlate with complementary foreground features</u>
Thus far, we have shown that simple skeleton parsings, feature counts and comparisons to object size provide ample shape description information for several classification scenarios with low computational effort and without user-defined input that may induce bias. While this achieves our original goal of analyzing shape in a dimensionality reduced setting and without the need for surface area approximation, we chose to push napari-toska further by combining foreground (*i.e.*, instance segmentations) analysis results with those of the background. This allowed us to both ensure the topological preservation of datasets during analysis and to confirm if object topology matches the assumed topology.

In many biology disciplines, it is generally assumed that a cell is sphere-like, meaning that is has a single bounding surface containing a continuous volume. In topology, 3D objects such as spheres and deformed spheres (e.g., an ellipsoid) have an ***Euler characteristic*** of 2. The Euler characteristic is an invariant feature (i.e., it will not change when an object is deformed or rotated) and therefore can be used in combination with other invariant features to consistently describe an object. In 2D, the topological equivalent of a sphere is a disk, which has an Euler characteristic of 1. In order to check an object's Euler characteristic using extracted topological skeletons, we took advantage of what is known as the ***cycle space*** in network science. In essence, a cycle is present in a network when a path exists that starts and ends with the same node without needing to traverse any edges more than once (Figure 6A). If the topological skeleton network of a cell's 2D instance segmentation contains a cycle, it implies that there is a hole in the instance segmentation (Figure 6A,B). Topologically, such a 2D instance segmentation would represent a circle-like object, which has an Euler characteristic of 0, rather than a disk-like object. We find that, while infrequent at the population level, these segmentation errors do exist and therefore the assumed object topology does not reflect the observed object topology (Figure 6A-C). The ability to detect these segmentation errors without user input is highly useful as it can be used to create a segmentation algorithm checkpoint where the instance segmentations for objects corresponding cycle-containing topological skeleton networks can be automatically re-evaluated and potentially corrected without time consuming user manual correction.

Mathematically, the foreground of an instance segmentation can be considered a ***set*** and it's background is the ***complement***. As the name suggests, a complement's behavior(s) should balance that of the set. In terms of topological skeletons, this means that in 2D datasets where the foreground contains disk-like objects the background's skeleton network should contain cycles. Precisely, the background's skeleton network should contain as many cycles as the foreground contains disk-like objects. If an object in the foreground (i.e., set) is not entirely spatially surrounded by the background (i.e., complement) a cycle will not emerge in the background's network. In the case of the 2D cell segmentations we have been examining, this is what we observe (Figure 6D). Also in line with the complement balancing observations of the set, we see that where instance segmentation errors are observed, an isolated sub-network emerges in the background skeleton network that reflects the shape of the error (Figure 6E). We can determine the degree of entirely contained foreground objects through the cycle-cell count ratio, which should equal one in a perfect dataset (Figure 6F). If the instance segmentation error were to be corrected in the foreground, the isolated sub-network would disappear (i.e., the pixels are correctly reassigned to belong to the set). This change would also be reflected in both the foreground and background Euler characteristics in order to preserve the dataset's topology. As an example, an image that is assumed to contain one disk-like object is revealed to be erroneously segmented if topological skeleton network analysis reveals the set's Euler characteristic to be 0 and the complement's Euler characteristic of 1. Upon segmentation correction, the set's Euler characteristic would become 1; the complement's, 0. By monitoring cycle-cell count ratios and changes in Euler characteristics to match an informed prediction, an algorithm could self-correct such errors.

Extending background skeleton observations to 3D requires the introduction of voids, like we see in semi-hollow spheres (Figure 4E, 6G, Movie S15). Paralleling the 2D relationship between cycles and foreground objects, in a 3D dataset's background skeleton, we should observe an equal number of voids to that of fully contained foreground objects. In scenarios where a 3D instance segmentations intersect with either the dataset's boundary or each other (Figure 5D), the expected voids will not emerge but cycles and/or branching may (Figure 6G,G'). As with 2D datasets, if the expected result is for all foreground objects to be separated, this could potentially be used a segmentation accuracy checkpoint. Detected voids also offer the opportunity to be used as a domain proxy. For instance, if an experimental scenario does not permit segmentation of cell membranes, voids detected in the background skeleton from a foreground nuclear segmentation can act as proxies for putative cell domains (Figure 6H). Effectively, such an approach allows an entire field of view to be dissected in a shape-informed fashion. This could be considered akin to methods like watershed. Combined, cycle and void detection in 3D offer ample opportunities to either refine/correct foreground segmentation results or further parse the background into relevant domains (Figure 6I).

## Discussion

We have shown that by combining digital topology, network metrics and spatial feature embedding it is possible to capture and simplify large amounts of shape information from seemingly convoluted volumes (*e.g.,* single mesenchymal cells). While many of the examples presented here are biological in nature, napari-toska is not limited by discipline or interest. This flexibility in application is further enhanced by the ease of use of the plug-in through its design of dropdown menus and recommended defaults (Movie S16-18). Of the several observations and trends presented in the previous sections, we would like to draw attention to the following three that we believe offer opportunities to advance topological skeletal analysis for use in instance segmentation shape description.

First, if the number of sides in a regular polygon increases while the polygon's area remains constant, its skeleton will eventually collapse to a single point due to pixel resolution limitation(s). One can imagine this process most simply as the discrete representation of a circle. This links the level to which object's shape can be described to the pixel/voxel resolution of a dataset. Similarly, with further analysis, it may be possible to collapse the polygon class circumscribing radius trends onto a single curve revealing a governing equation for polygons. Potentially, determining how distinct reference shapes (*e.g.*, polygons) can be detected at a certain pixel/voxel resolution could be used to determine if it is ideal for a proposed imaging experiment. For example, if an experiment needs to detect fine membrane projections or invaginations but the current pixel resolution cannot differentiate between an octagon and a decagon, the user would know to increase the resolution. Further analysis of reference shape topological skeleton behavior(s) is required to determine any absolute limits that may exist for the method. It would also be necessary to curate several example datasets from popular sample types, imaging modalities and resultant segmentations to create a proof of concept for this to come to fruition; however, if undertaken, the results would be a valuable future resource for experimental design.

In an adjacent vein, the apparent power law behavior observed between skeletal size and branching indicate that topological skeletons could be used to evaluate surface roughness. Assuming the appropriate resolution at which to acquire a dataset has been determined, subsequent assessment of surface roughness can inform either on the performance of an instance segmentation algorithm or on an object's condition (*e.g.*, experimental phenotype). The observations of periodic branching behavior can aid here. We briefly and superficially showed that digital rotation of objects, be they simple convex or complex with concavities, exhibit periodic emergent branching behaviors (Movie S8). This can aid in realizing the limits of surface roughness characterization due to discrete space if the process of excessive branching due to false vertex detection can be analytically inverted. If achieved, this will allow the reliable creation of rotationally invariant topological skeletons thereby improving surface variation representation. However, this will require further investigation, and, critically, a the curation of segmentations from well-characterized systems with multiple conditions exhibiting varied surface roughness. This can be complemented by an extended evaluation of surface concavity-containing reference shapes in both 2D and 3D, which could also then be used as shape basis for positive/negative controls to experimental datasets.

Finally, napari-toska's utilization of cycle/void detection in both instance segmentation foreground and background offers several directions for method advancement. napari-toska's capability to automatically detect foreground instance segmentation errors is highly useful. We have begun to extend napari-toska's functionality to include proposed segmentation error corrections, such as the errors shown in Figure 6. As this will be an automated method, users will not have to painstakingly manually correct segmentations, which can be incredibly time consuming. Interestingly, for the 2D datasets tested for cycle detection, we see that if a cycle emerges in an object's network it is more likely to reoccur at a later timepoint (data not shown). It may be significant to test if this is true across multiple segmentation algorithms to determine if there is an underlying and thus far undetected bias to retain externally introduced artifacts.

*In toto* background skeletons, meaning all branches, cycles, voids and isolated subnetworks, could be used to ask questions of foreground object domain. A background skeleton void could be directly used as a domain proxy in place of a foreground membrane segmentation. It could also be used as a comparison to the results of common segmentation methods (*e.g.*, distance transform followed by watershed). In such cases, background skeleton networks can act as the foundation to ask where and how efficiently signals could be distributed across a dataset's full volume depending on where the source(s) are in the foreground. For example, one could examine all possible paths for a diffusive signal molecule to travel through a background skeleton network. If the background network edges were weighted by some functions coupling diffusion kinetics to foreground source node proximity, this would significantly accelerate *in silico* experiments by refining the effective variable range(s) governing the phase space of the system (*i.e.*, signal reception is position A results in shape phenotype X, while signal reception in positions A and B results in shape phenotype B). One could imagine other systems with similar applicability such as bacterial (foreground) locomotion through regions of low physical resistance (background). However, the ability to perform these *in silico* experiments requires determining the minimal set of governing rules for how foreground and background networks can communicate. Because cross-network communication is still a largely open question in many fields, this necessitates significant method expansion and collaborative effort. Initial steps are currently being taken towards this goal and we hope that shape behaviors revealed here by napari-toska will serve as a solid foundation.

## Study Limitations

The datasets used in this manuscript to show napari-toska's capabilities have largely been biological samples existing in the micrometer to millimeter range. However, the plug-in's functionalities are agnostic to range. Currently, napari-toska works creates and analyzes only undirected network graphs. This may be subject to change in the future depending on community usage and the research directions of the authors.

## Author Contributions

AQR and CM designed the project. AQR and CM inspected the relationships between topological sub-disciplines. AQR wrote the original python code. AQR and JS implemented the code as a napari plug-in. AQR, JS and CM composed the manuscript.

## Acknowledgments

We thank Jacqueline M. Tabler, Diana Alves-Afonso, Adrian Lahola Chomiak, Robert Haase and the Christoph Zechner group for discussions through the development and implementation of the project. We thank Markus Mukenhirn, Tiger Lao and Deepika Sundarraman for manuscript proofreading.

## Declaration of Interests

The authors declare no competing interests.

## Figure Titles & Legends

**Figure 1. Digital topology and network science principles for graph representation of topological skeletons.** (A) Object representation of the napari-toska logo in continuous and discrete space. Local over/under representation in discrete space compared to continuous (*i.e.*, true representation) highlighted. (B) Schematic representations of neighbor defining relationships in 2D and 3D discretization. Green indicates a valid vertex, edge or face neighbor. (C) Perspective views of kernels used for evaluation of neighborhoods in 2D and 3D datasets where the central pixel/voxel of the kernel is overlapped with the pixel/voxel being evaluated. (D) Schematic examples of end, chain and branch point classification scenarios in 2D and 3D discrete neighborhoods. (E) Conversion schematic of the discretized napari-toska logo's topological skeleton into a network highlighting the endpoints (turquoise), branch points (orange), edges(black/purple) and the network's spine (purple).

Glossary terms: branch, branch point, chain point, edge, endpoint, face, kernel, node, neighborhood, $N_4$, $N_8$, $N_6$, $N_{18}$, $N_{26}$, topological skeleton, undirected network graph, weighted network graph.

**Figure 2. Emergent topological skeleton elongation and branching behaviors in polygons and simple curved objects in 2D.** (A) Representative regular polygons of Class 1 (square), Class 2 (hexagon) and Class 3 (equilateral triangle) with topological skeletons overlaid. (B) Line plot of the circumscribing circle's radius (x-axis) for Class 1 ($n_{sides}$ = 4,8,12,16,20), Class 2 ($n_{sides}$ = 6,10,14,18) and Class 3 ($n_{sides}$ = 3,5,7,9,11,13,15,17,19) regular polygons relative to the ratio of topological skeleton size in pixels to polygon side length (y-axis). Mean ($\mu$), solid purple, turquoise and orange lines. 1 standard deviation ($\sigma$), shaded purple, turquoise and orange areas. (C) Line plot showing the rotation angle of an irregular 4-sided polygon against the number of branches in its skeleton. $5^{o}$ (turquoise), $45^{o}$ (purple) and $90^{o}$ (orange) angles showing robust periodic branching events highlighted. (D) Ellipses with topological skeletons overlaid. Top, minor-major axis length ratio = 0.6. Bottom, minor-major axis length ratio = 0.2. (E) Scatter plot showing the linear relationship between ellipse minor-major axis length ratio (x-axis) to ellipse skeleton-major axis length ratio (y-axis). Axis aspect ratio values of 0.2, 0.4, 0.6 and 0.8 are highlighted. See STAR Methods for axis metric calculation. (F) Crescent moon-like shape with topological skeleton overlaid. Sphere 1-Sphere 2 mismatch = 0.2. (G) Pac-Man-like shape with topological skeleton overlaid. Sphere 1-Sphere 2 mismatch = 0.2. (H) Scatter plot showing the relationships of crescent moon-like and Pac-Man-like shape generation ratios (x-axis; see Methods and Glossary Appendix for detailed definitons) relative to the skeleton size-major radius of curvature (y-axis). Radial offset ratio values of 0.2, 0.4, 0.6 and 0.8 are highlighted. Shape generation ratio range where Pac-Man-like skeletons posses branching events is shaded in gray. See STAR Methods for axis metric calculation.

Glossary terms: branch, class 1 regular polygon, class 2 regular polygon, class 3 regular polygon, regular polygon, topological skeleton.

**Figure 3. Unique object and population shape dynamics are tractable through topological skeleton network conversion and analysis.** (A) Single timepoint image of a 2D fluorescence imaging timelapse dataset of motile, non-confluent mesenchymal cells expressing a membrane marker (grayscale lookup table). (B) Labeled instance segmentations corresponding to the fluorescence image in (A) (white-hsv lookup table). (C) Topological skeleton extraction corresponding to the instance segmentations shown in (B) (white-hsv lookup table). (D) Spines corresponding to the topological skeletons shown in (C) (white-hsv lookup table). (E) Line plots showing topological skeleton features for a single cell from dataset shown in (A-D) over time. (F-K) Line plot showing the population mean ($\mu$) and standard deviation ($\sigma$) of the endpoint count (F), branch point count (G), branch count (H), branch length (I, in $\mu$m), and spine lengths (J, in $\mu$m) over time of all cells segmented in the experiment shown in (A-D). (K) Schematic representation of an average cell's topological skeleton network based on population statistics over the entire timelapse.

Glossary terms: branch, branch point, endpoint, instance segmentation, set, topological skeleton, undirected network graph, weighted network graph.

**Figure 4. 3D object skeletonization shows parallels to 2D elongation behaviors, while surface concavities induce more complex branching.** (A) Perspective 3D view of an ellipsoid with major-minor(s) axis length ratio = 0.6. (A') Central slice view through the ellipsoid shown in (A) overlaid with its topological skeleton. (B) Scatter plot showing ellipsoid $1^{o}$ minor-major axis length ratio (x-axis), $2^{o}$ minor-major axis length ratio (datapoint color index) relative to skeleton length (y-axis, normalized to maximum skeleton length for all ellipsoids analyzed). (C) Perspective 3D view of a budding yeast-like reference shape with a bud-mother sphere radius length ratio = 0.6. See STAR Methods for axis metric calculation. (C') Central slice of budding yeast shown in (C) overlaid with its topological skeleton. (D) Scatter plot showing the exponential relationship of bud-mother sphere radius ratio (x-axis) relative to skeletal size (y-axis; normalized to object volume). (E) Perspective 3D view of a sphere ($s_1$) with a prominent surface concavity ($c_1$) with radius = $0.4r_{s1}$ and origin point with distance $0.98r_{s1}$ from the origin point of $s_1$. See STAR Methods for axis metric

calculation. (E′) Central slice through volume-skeleton overlay of sphere ($s_1$) with branching inducing surface concavity (c) with radius = $0.4r_{s1}$ and origin point $0.66r_{s1}$ from the origin point of $s_1$. (E′′) Central slice through volume-skeleton overlay of a semi-hollow sphere where the void has radius = $0.4r_{s1}$ and origin point with distance $0.56r_{s1}$. (E′′′) Central slice through semi-hollow sphere ($s_1$) with centralized void with radius = $0.4r_{s1}$. (F) Scatter plot showing the relationship between the shape generation ratio of semi-hollow spheres or spheres with significant surface concavities (x-axis) with either normalized skeleton length in voxels or normalized branch count (organ or turquoise, respectively; y-axis). See STAR Methods for axis metric calculation. (G) Scatter plot showing the relationship between branch count (x-axis) with normalized skeleton size in voxel count (y-axis) for multiple shape generation ratios of intersecting spheres. Line plot showing collapse onto a potential power law fit over one order of magnitude (green dashed line; $y = 0.867x – 2.840$).

Glossary terms: branch, topological skeleton.

**Figure 5. Topological skeleton analysis can distinguish experimental phenotypes and shape features of cellular compartments.** (A) Single timepoint maximum z-projection of fluorescence membrane signal instance segmentation for a single timepoint from an experimental imaging dataset of an individual cell embedded in a 3D Matrigel matrix. Scale bars = $10\mu$m. (A′) Kamada-Kawai network representation (Kamada and Kawai, 1989; Mingers and Leydesdorff, 2015; Aria and Cuccurollo, 2017) of the extracted topological skeleton for the segmentation shown in (A) with the spine contributing edges and nodes highlighted in purple. (B) Single timepoint maximum z-projection of pseudo-fluorescence membrane signal instance segmentation for a single timepoint from a simulated imaging dataset of an individual cell embedded in a 3D Matrigel matrix. Scale bars = $10\mu$m. (B′) Kamada-Kawai network representation of the extracted topological skeleton for the simulated segmentation shown in (B) with the spine contributing edges and nodes highlighted in purple. (C) Line plots showing the temporal dynamics of normalized spine length and end-branch point count ratios for the experimental (solid lines) and simulated (dashed lines) datasets shown in (A,B). Each category's spine length is normalized to its corresponding timelapse's minimal length measurement. (D) Sum projection over the z-axis of a 3D nuclear instance segmentation. (D′) Perspective projection of a single nucleus' topological skeleton from (D) inset marked by green dashed line.. (E) Sum projection over the z-axis of a 3D membrane instance segmentation. (E′) Perspective projection of a single nucleus' topological skeleton from (E) inset marked by green dashed line. (F) Line plot comparing endpoint-branch count ratio mean ($\mu$) and standard deviation ($\sigma$) temporal dynamics of nuclear and membrane topological skeleton networks.

Glossary terms: branch point, endpoint, instance segmentation, Kamada-Kawai layout, set, undirected network graph, weighted network graph.

**Figure 6. Background skeletonization, cycle detection and isolated sub-network detection can identify segmentation errors.** (A) Cycle presence detected in a single cell's topological skeleton in four timepoints that correlate to errors in the cell's instance segmentation (bottom). Network representations (top) are drawn in Kamada-Kawai layout with spine edge and node components highlighted in purple. (B) Schematic representation showing how cycle emergence alters the average cell's topological skeleton network based on population statistics from Figure 3. (C) Line plot showing the mean ($\mu$) and standard deviation ($\sigma$) of population level cycle emergence count in cell instance segmentations over time in the dataset corresponding to (A), (B) and Figure 3. (D) Single timepoint overlay showing foreground instance segmentations of cells (hsv lookup table) with cycle containing background topological skeleton (black). (E) Insets of (D) timelapse showing the same cell and timepoints as (A). Isolated background skeleton subnetworks can be seen filling the areas of foreground segmentation errors. (F) Line plot showing the population level mean ($\mu$) and standard deviation ($\sigma$) of foreground cell-background cycle count ratio over time in four datasets. (G) Sum projection over the z-axis of a 3D instance segmentation's background topological skeleton. (G′) Perspective projection of green-dashed line inset highlighted in (G) showing multiple cycles corresponding where a single nucleus' instance segmentation (see Figure 5D) is not fully contained by the image volume's boundaries. (H) Sum projection over the z-axis of voids detected in the background skeleton of a 3D instance segmentation. (I) Line plot comparing feature counts detected in either 3D foreground instance segmentation (nuclei) or corresponding background skeleton (cycles and voids) dataset.

Glossary terms: complement, cycle space, Euler characteristic, instance segmentation, Kamada-Kawai layout, set, topological skeleton, undirected network graph, weighted network graph.

## STAR Methods

### Resource Availability

The plugin and its installation instructions will be publicly available upon publication and can be found in the following GitHub repository: https://github.com/allysonryan/napari-toska. Similarly, all original code relevant to the creation of the plugin as well as original code for the generation of reference shapes relevant to reproducing results as well as Figures 1, 2 and 5 will be publicly available in the following GitHub repositories: https://github.com/allysonryan/reference-shape-generation, https://github.com/allysonryan/toska. The datasets used for results relevant to Figure 3, 4 and 6 are publicly accessible through the following link(s): https://celltrackingchallenge.net/ (Maška et al., 2014; Ulman et al., 2017). GitHub repositories will be activated on Zenodo, relevant DOIs generated and added to the Key Resources Table upon publication. Any additional information required to reanalyze the data reported in this paper is available from the lead contact, Carl D. Modes (modes@mpi-cbg.de), upon request.

## STAR Methods

### Resource Availability

The plugin and its installation instructions will be publicly available upon publication and can be found in the following GitHub repository: https://github.com/allysonryan/napari-toska. Similarly, all original code relevant to the creation of the plugin as well as original code for the generation of reference shapes relevant to reproducing results as well as Figures 2, and 4 will be publicly available in the following GitHub repositories: https://github.com/allysonryan/reference-shape-generation, https://github.com/allysonryan/toska. The datasets used for results relevant to Figure 3, 5 and 6 are publicly accessible through the following link(s): https://celltrackingchallenge.net/ (Maška et al., 2014; Ulman et al., 2017). GitHub repositories will be activated on Zenodo, relevant DOIs generated and added to the Key Resources Table upon publication. Any additional information required to reanalyze the data reported in this paper is available from the lead contact, Carl D. Modes (modes@mpi-cbg.de), upon request.

### Method Details

**Reference shape generation.** To create a crescent moon-like shape, two circles with equal radii are perfectly overlapped, then one circle's origin point is shifted uni-axially by some distance and the resultant exclusions create identical crescents. We retain one and generate its skeleton (Figure 2F). The origin shift is referred to as the primary($1^{\circ}$)-secondary($2^{\circ}$) origin ratio (Figure 2H). Pac-Man-like shapes are generated by placing the origin point of a circle with radius $kx$ on the perimeter of a circle with radius $x$. The larger exclusion area remaining from the larger circle results in a Pac-Man-like shape (Figure 2G). $k$ is referred to as the primary($1^{\circ}$)-secondary($2^{\circ}$) radius ratio (Figure 2H).

**Normalization equations and plot axis ratios.** Figure 2B y-axis ratio is the sum of pixels (xy-resolution=1) in the regular polygon's topological skeleton divided by the length of a single side of the regular polygon in continuous space. Figure 2E y-axis ratio is the sum of pixels (xy-resolution=1) in the ellipse's topological skeleton divided by the length of it's major axis in continuous space. Figure 2H x-axis ratio is the shape generation ratio of each corresponding shape (crescent or Pac-Man, see Reference Shape Generation). Figure 2H y-axis ratio is the sum of pixels (xy-resolution) in the object's topological skeleton divided by its primary (1$^{o}$) circle's radius in continuous space. Figure 4B x-axis is the length of an ellipsoid's primary (1$^{o}$) short axis length divided by its major axis length, both in continuous space. Figure 4B's y-axis is the sum of voxels (xyz-resolution=1) in the ellipsoid's topological skeleton divided by the maximal sum of voxels for any topological skeleton in the same category, which is defined by the x-axis value. Figure 4D x-axis ratio is the length of a bud sphere's radius divided by the larger mother sphere's radius in continuous space. Figure 4D y-axis is the sum of voxels (xyz-resolution=1) in a budding yeast's topological skeleton divided by the sum of voxels (xyz-resolution=1) in the budding yeast's volume. Figure 4F x-axis is the 3D generalization of the 2D Pac-Man shape generation ratio (see Reference Shape Generation). Figure 4F y-axis is either the semi-hollow sphere branch count or skeleton length of a, which equals voxel (xyz-resolution=1) sum divided by the respective feature's maximum in the same category, which is defined by the x-axis value. Figure 4G y-axis is the size of a semi-hollow sphere's topological skeleton's (equal to the sum of its voxels with xyz-resolution=1) divided by the maximum size per category, which is defined by its shape generation ratio.

**Code.** All original code resulting from this project was developed in Python using functions from itertools, networkx, numpy, os, scipy, and skimage packages. The plugin was created using cookiecutter. Figure image panels were prepared with Fiji and Python (Schindelin et al., 2012). Figure network and plot panels were prepared using Python's networkx and matplotlib.

Quantification and Statistical Analysis

All statistical calculations and fits were performed in Python. The means ($\mu$) and standard deviations ($\sigma$) for results relevant to Figures 2,3 and 6 were calculated using Python numpy functions. Plot log scales and normalizations were performed using matplotlib. The power law fit for Figure 4G was predicted using sklearn.linear_model.

## Supplemental Information Titles & Legends

**Appendix 1. napari-toska glossary.** Definitions of digital topology, network science and image analysis terminology relevant to this study.

**Movie S1.** Class 1 regular polygons with an increasing number of sides.

**Movie S2.** Class 2 regular polygons with an increasing number of sides.

**Movie S3.** Class 3 regular polygons with an increasing number of sides.

**Movie S4.** Rotational effect on Class 1 regular polygon skeletal lengthening and branching. Top to bottom, polygon number of sides = 4,8,12,16,20. Left to right, polygon, topological skeleton, spine.

**Movie S5.** Rotational effect on Class 2 regular polygon skeletal lengthening and branching. Top to bottom, polygon number of sides = 6,10,14,18. Left to right, polygon, topological skeleton, spine.

**Movie S6.** Rotational effect on Class 3 regular polygon skeletal lengthening and branching. Top to bottom, polygon number of sides = 3,5,7,9,11,13,15,17,19. Left to right, polygon, topological skeleton, spine.

**Movie S7.** Rotational effect on 4-sided irregular polygon skeletal lengthening and branching.

**Movie S8.** Rotational effect on curved objects with increasing complexity – kidney bean (top), hands (middle) and maple leaf (bottom).

**Movie S9.** Ellipse major-minor axis length ratio effect on skeletal lengthening (ellipse area, gray; skeleton; black). Ratio stamped on each frame.
**Movie S10.** Shape generation ratio effect on skeletal lengthening in crescent moon-like shapes (crescent moon-like area, gray; skeleton; black). Ratio stamped on each frame.
**Movie S11.** Shape generation ratio effect on skeletal lengthening and branching in Pac-Man-like shapes (Pac-Man-like shape area, gray; skeleton; black). Ratio is stamped on each frame.
**Movie S12.** Perspective projection and 360° revolution of ellipsoid with minor-major axis ratio =0.6.
**Movie S13.** Perspective projection and 360° revolution of budding yeast with bud-mother sphere radius ratio = 0.6.
**Movie S14.** Perspective projection and 360° revolution of death star with concavity-sphere radius ratio = 0.4.
**Movie S15.** Perspective projection of 3D nuclear segmentation background skeleton over time.
**Movie S16. napari-toska extracts and parses complex skeletons of arbitrary 2D objects.** 2D labeled skeletonization and skeleton parsing of all objects in the field followed by branch labeling and spine extraction of a single object of interest.
**Movie S17. napari-toska extracts and parses complex skeletons of arbitrary 3D objects.** 3D labeled skeletonization and skeleton parsing of a simple sphere and a complex, loop-containing branched structure.
**Movie S18. napari-toska pipelines can be designed and executed through dropdown menus and/or default settings.** Screen recording of basic pipeline design, execution, quantitative analysis output and interactions between results table outputs and skeleton image visualizations (*i.e.*, LUTs).

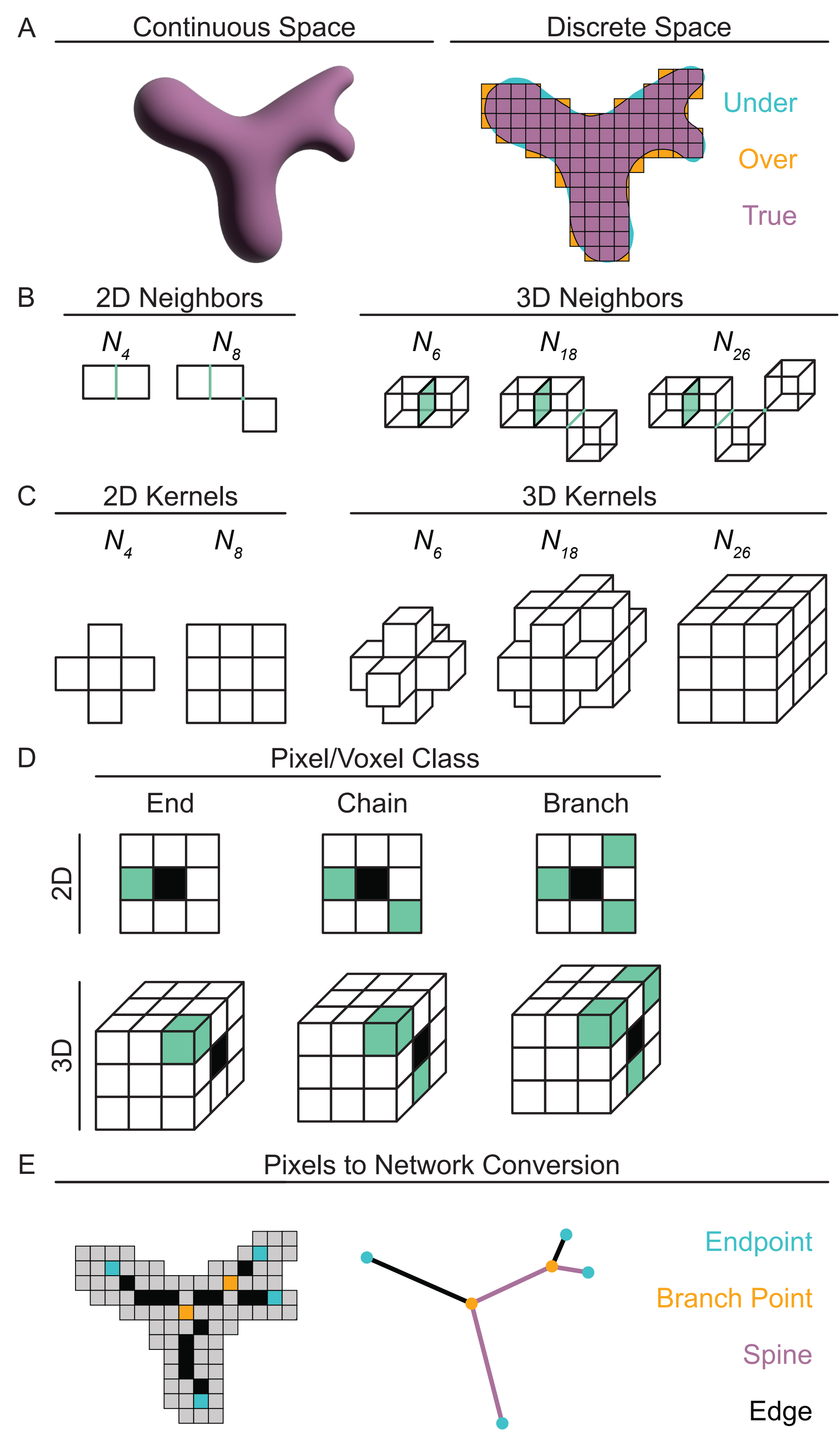

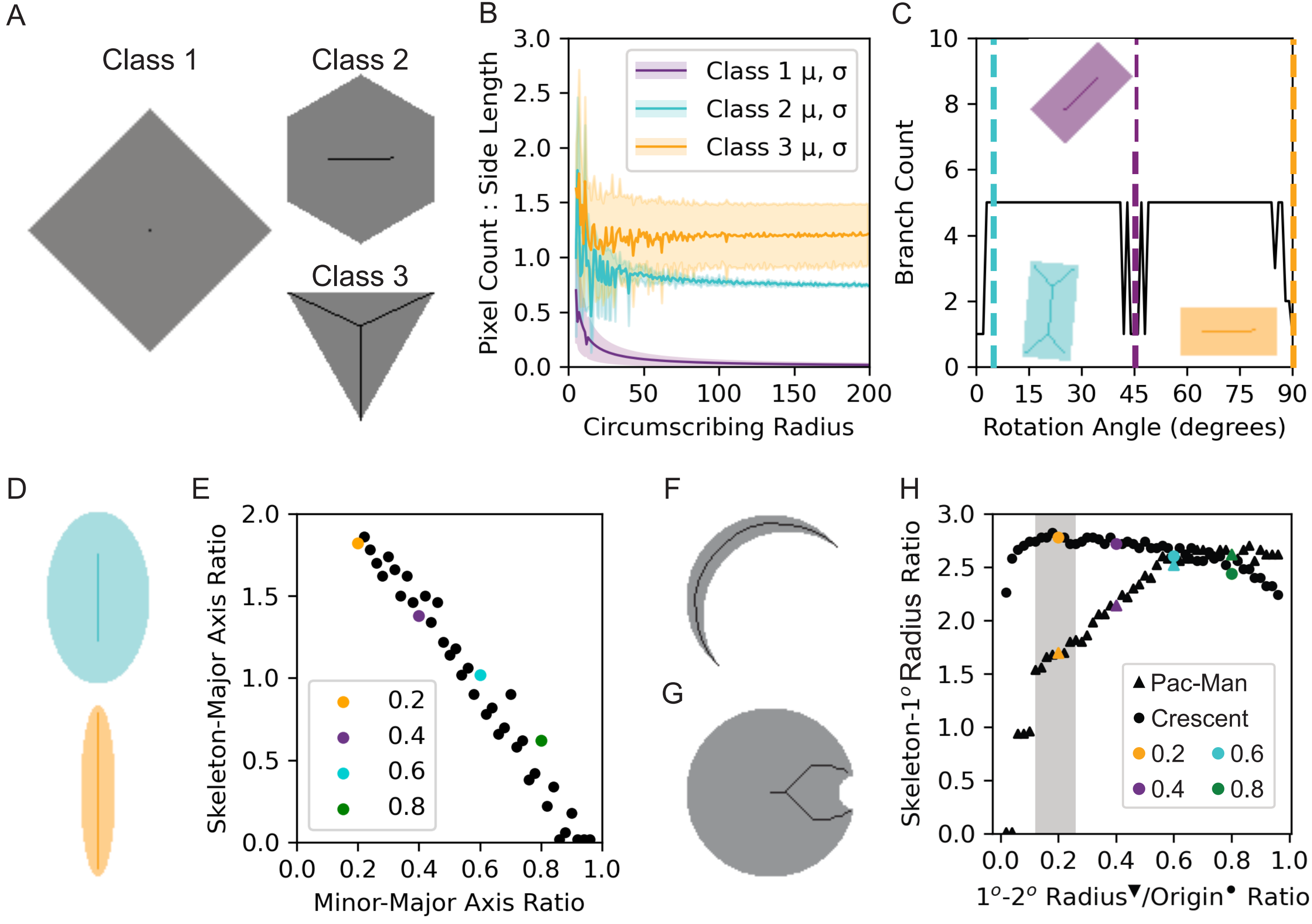

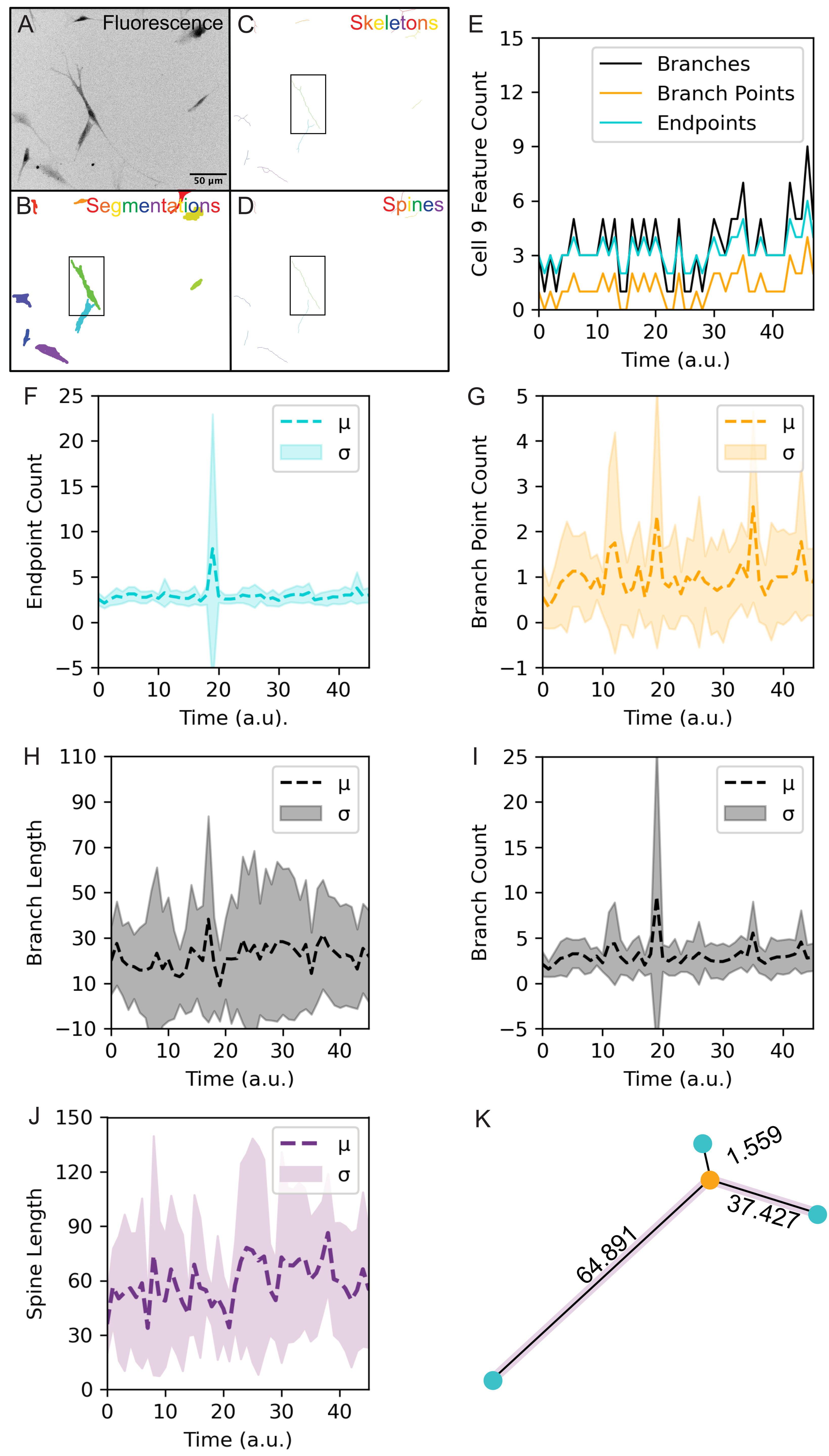

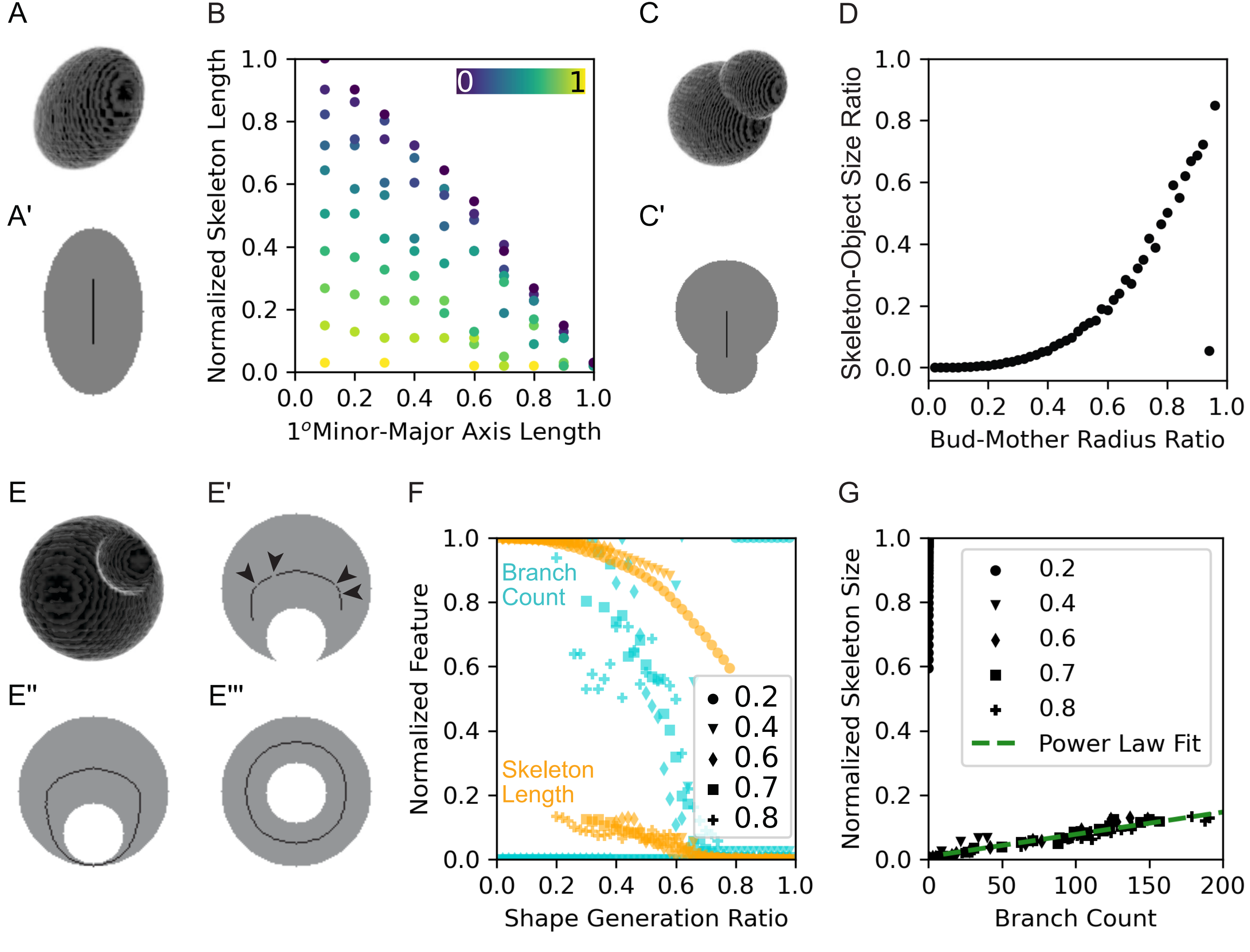

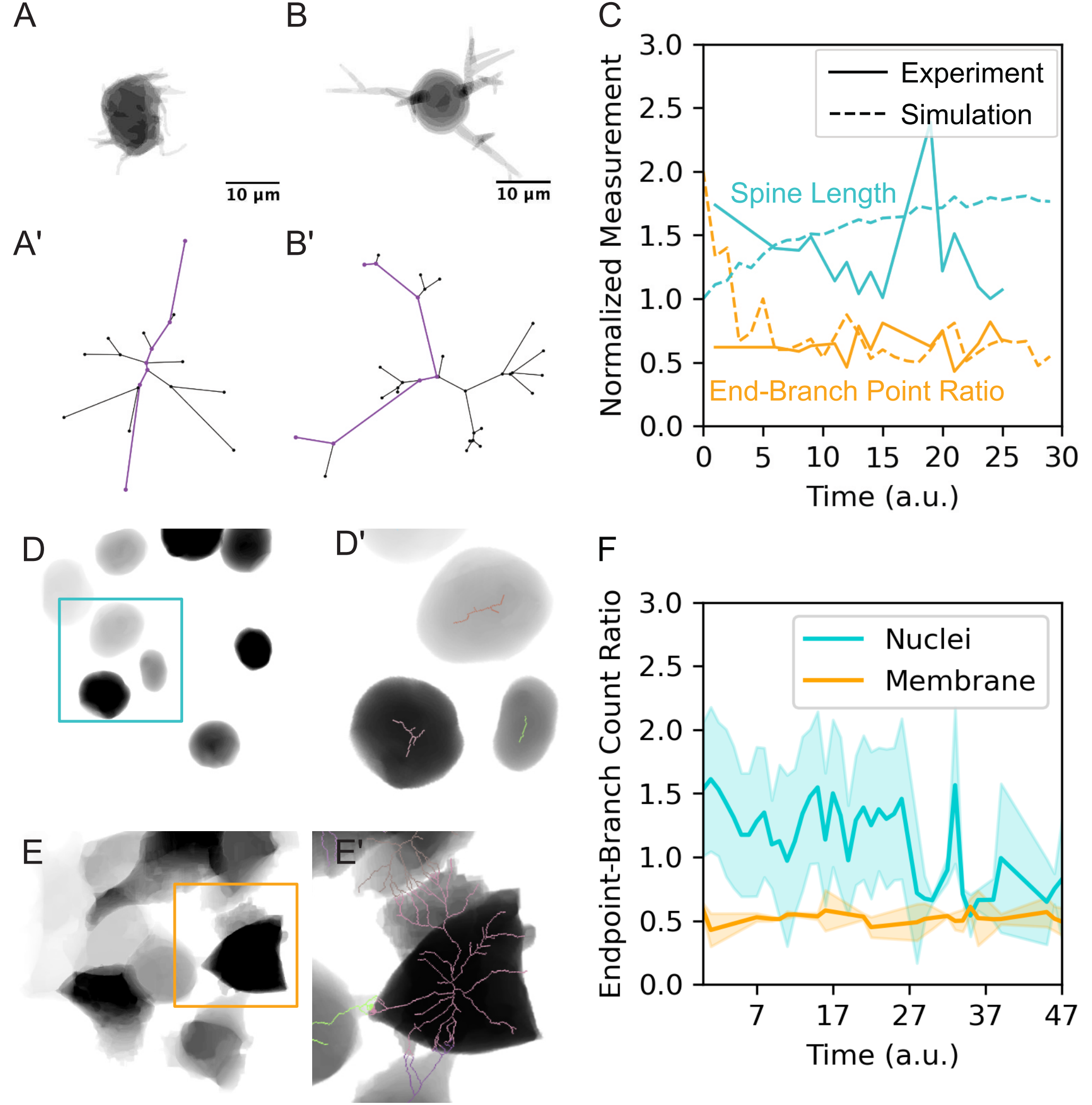

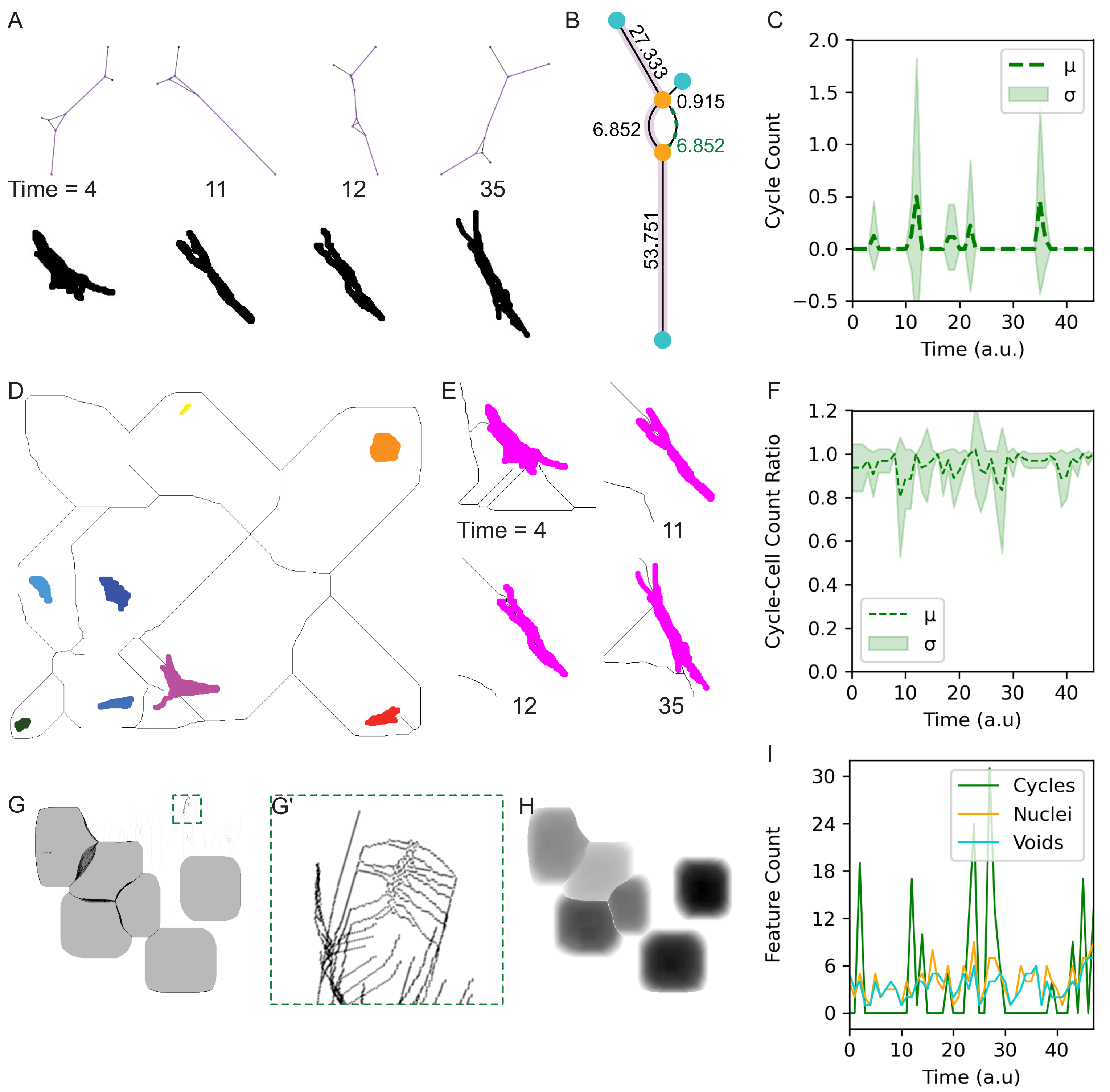

## Appendix 1: napari-toska Glossary

- Branch: a collection of connected chain pixels/voxels in which each pixel/voxel's neighborhood sum equals 2. Two of the pixels/voxels in a branch have a neighboring pixel/voxel that is classified as either an endpoint or branch point pixel/voxel.

  Relevant Figure Panel(s). Figure 1D, 1E, 2C, 3E, 3H, 3I, 4F and 4G.

- Branch point: a pixel/voxel where the sum of its neighbors is equal to or greater than 3.

  Relevant Figure Panel(s). Figure 1D, 1E, 3E, 3G, 5C and 5F.

- Chain point: a pixel/voxel where the sum of its neighbors is equal to or greater than 2. A collection of adjacent chain points creates a branch (Figure 1D).

- Class 1 regular polygon: a polygon where the number of vertices is divisible by 4 and all sides are of equal length.

  Relevant Figure Panel(s). Figure 2A-C, Movie S1, S4 and S7.

- Class 2 regular polygon: a polygon where the number of vertices is divisible by 2 but not 4 and all sides are of equal length.

  Relevant Figure Panel(s). Figure 2A-C, Movie S2 and S5.

- Class 3 regular polygon: a polygon where the number of vertices is not divisible by 2 and all sides are of equal length.

  Relevant Figure Panel(s). Figure 2A-C, Movie S3 and S6.

- Complement: the set of all items that are not in the set such that the union of the set and the complement is equal to an entire dataset (*i.e.*, the universal set).

  Example(s). In a headshot with a solid color backdrop, the pixels showing the person is the set, and the pixels showing the backdrop are the complement. In image instance segmentations, the set and its complement can also be thought of as the foreground and background, respectively.

  Relevant Figure Panel(s). Figure 6D-I.

- Continuous space: a system where space and/or time are infinitely divisible. The base unit in such a system is a point. Transitions between points in the system are smooth.

- Cycle space: the vector space composed of all possible cycles in a network. Cycle space is equivalent/isomorphic to the first homology group ($H_1$).

  Example(s). The simplest cycle containing network is a triangle, which has 6 elements (3 nodes and 3 edges). This network's cycle space contains all of these elements because all are required to complete the one cycle (*i.e.*, loop).

  Relevant Figure Panel(s). Figure 6A-I.

- Digital topology: a sub-discipline of topology concerned with understanding how topological features, such as boundary surfaces, translate and are preserved when translating objects from continuous space to digital representations (*e.g.*, images). A well known application of digital topology is connected component analysis.

- Directed graph: a network in which edges can only be traversed in a specific direction (e.g., you can move from node A to node B over edge 1, but you cannot move from node B to node A over edge 1).

  Example. In a simple triangular network graph with nodes $A, B, C$, and only three edges where each edge can only be traversed in a single direction, one possible arrangement of edge directions is $A \rightarrow B,\ B \rightarrow C,\ C \rightarrow A$. There are several other possible configurations.

- Discrete space: a system where space and/or time are not infinitely divisible. Rather than exhibiting smooth transitions between radii of curvature, changes in a line or boundary's curvature can only be assessed at defined step sizes.

  Example(s). Images, such as microscopy datasets. The precision of step sizes are determined by the pixel/voxel resolution.

- Distance transform: a function that represents every pixel/voxel of a discrete dataset relative to its spatial distance to a particular pixel(s)/voxel(s) of reference. In image processing, this is often coupled with a watershed transform and connected component analysis to create an instance segmentation.

  Example(s). Consider a binary image containing a single circle. Applying a distance transform function changes the value each foreground (set) pixel's value from 1 to the distance of that pixel from the circle's origin (*i.e.*, its center of mass).

- Edge: in discrete space, the bounding lines of a pixel/voxel that exist between two vertices. In network science, an edge is the base unit of connection between two nodes or a node with itself.

Example(s). In an image, a pixel has four edges. In the simplest network containing a cycle but no self-loops, there exist three edges (*i.e.*, a triangle).

Relevant Figure Panel(s). Figure 1E.

- Edge weight: an attribute assigned to an edge to reflect the likelihood that it will be visiting in a network path. Weights can be either stable (*i.e.*, assigned as an unchanging number) or dynamic (*i.e.*, defined by a single or multi-variate function).

- Endpoint: in images, a pixel/voxel where the sum of its neighbors is equal to 1. When translating images into networks, an endpoint can be thought of as a node with degree equal to 1.

Relevant Figure Panel(s). Figure 1D, 1E, 3E, 3F, 5C and 5F.

- Euler characteristic: a shape descriptor from the field of algebraic topology. Because it is topologically invariant, the Euler characteristic for a perfect circle/sphere is the same as for any bent or deformed circle/sphere as long as the bounding perimeter(s)/surface(s) are not disrupted, fragmented or merged. This makes the Euler characteristic an ideal metric to ensure topological preservation in both discrete and continuous space datasets.

Example(s). For all regular convex polyhedra, the Euler characteristic is 2. This is determined from the formula $\chi = V - E + F$ where $\chi$ is the Euler characteristic, $V$ is the number of vertices, $E$ is the number of edges and $F$ is the number of faces. If you semi-discretized the surface of a torus into a polygonal mesh, you can apply the same formula for an Euler characteristic of 0.

Relevant Figure Panel(s). Figure 6A-I and Movie S15.

- Face: in discrete space, the area of a voxel that is bounded by four vertices and four corresponding edges that form a bounded square existing in either a xy-,xz- or yz-axis parallel plane.

Example(s). In 2D images, each pixel of an image has one face. In 3D images, each voxel has six faces. The face's area is determined by the relevant xy-, xz- or yz-resolution.

Relevant Figure Panel(s). Figure 1B-D.

- Feature embedding: the process of retaining relevant system/object identifying information from higher dimensions during dimensionality reduction.

- Global neighborhood principle(s): rules that apply to every pixel/voxel of a dataset (e.g., a voxel shares its faces with six other voxels).

- Instance segmentation: the result of a pixel/voxel classification algorithm where each pixel of an input image is assigned to belong to a certain object. Each object is given a unique identifier. This is in contrast to semantic segmentation where pixels are classified by category and therefore pixels/voxels that are not connected may have the same identifier.

  Example(s). In a microscopy image of cells where a nuclear protein is tagged with a fluorescent protein, a high quality instance segmentation result contains the same number of unique identifiers as the original image does nuclei and the segmentation's foreground pixels correlate to the fluorescent signal.

  Relevant Figure Panel(s). Figure 3B, 5A, 5B, 5D-E', 6A, 6D and 6E.

- Kamada-Kawai layout: a force-directed network graph visualization algorithm. Imagine the connection between two nodes of a graph to be connected by a spring with an ideal length. The most desirable layout of all the nodes minimizes the summed energy of the springs in the system. The algorithm iterates over potential layouts of the nodes and finds the one with the minimal energy.

  Example(s). A square based pyramid can be represented by a network with 5 nodes and 8 edges. Each node has a degree of 3. Each edge has the same ideal length. The lowest energy planar visualization compresses 4 edges and stretches 4 edges. See Kamada & Kawai, 1989, Mingers & Leydesdorff, 2015 and Aria & Cuccurollo, 2017 for more detailed explanations and prior use cases.

  Relevant Figure Panel(s). Figure 5A', 5B' and 6A.

- Kernel: a matrix of specific size and shape with all element values defined. When iterated over an input dataset's elements, the kernel performs a convolution (*i.e.*, function) to render a certain output effect.

  Example(s). A common image processing kernel is an edge detector. In 2D, this can be have a cross shape when using the $N_4$ neighborhood, or a square share when using the $N_8$ neighborhood:

$$N_4 : \begin{bmatrix} 0 & -1 & 0 \\ -1 & 2 & -1 \\ 0 & -1 & 0 \end{bmatrix} \qquad N_8 : \begin{bmatrix} -1 & -1 & -1 \\ -1 & 2 & -1 \\ -1 & -1 & -1 \end{bmatrix}$$

  Relevant Figure Panel(s). Figure 1C.

- Line: a 1D entity in continuous space that can be infinitely subset into line segments, which are defined by points along the line between two bounding points.

- Medial Axis Transform (MAT): a shape description methods that renders the continuous space version of a topological skeleton. The MAT reduces a shape to a set of points and/or line segments representing all asymmetries. Variations of the MAT exist. It should be possible to fully recreate the original shape from the MAT without additional information.

    Example(s). See Blum 1967 and Blum 1973 for the original applications and results of the MAT.

- Minor-major axis length ratio: the minor axis length is equal to the minimal straight line distance from the object's origin to its boundary, and the major axis length is equal to maximum straight line distance from the object's origin to its boundary. The ratio of these two is often used to describe ellipses/ellipsoids and reflects elongation, or the deviation of an ellipse/ellipsoid from a perfect circle/sphere. This ratio is also commonly referred to as the aspect ratio.

- Node: the base unit of a network. A node represents a unique entity in the system of interest.

    Example(s). In a social network, each person is a node. In a microscopy image containing multiple cells, each cell can be considered as a node. When converting a topological skeleton into a network, each branching point and each endpoint can be considered a node.

    Relevant Figure Panel(s). Figure 1D-E.

- Network science: the study of connections between individual entities by representing each entity as a node and its connection with other entities through nodes. Classically, network science does not require nodes to be rigidly, positionally assigned. However, spatial features of entities and their relationships can be encoded through node/edge weights. Importantly, this makes network science amenable to application at any scale or across multiple scales.

- Neighborhood: the set formed by a pixel/voxel of interest and its possible adjacent pixels/voxels.

    Example(s). See definitions for $N_4$, $N_8$, $N_6$, $N_{18}$ and $N_{26}$.

    Relevant Figure Panel(s). Figure 1B.

- $N_4$: the minimal neighborhood in 2D, which is composed only of the pixel of interest and the 4 pixels that share an edge with it.

    Relevant Figure Panel(s). Figure 1B.

- $N_8$: the maximum neighborhood in 2D, which is composed of the pixel of interest, the 4 pixels that share an edge with it and the 4 pixels that share a vertex with it.

  Relevant Figure Panel(s). Figure 1B.

- $N_6$: the minimal neighborhood in 3D, which is composed only of the voxel of interest and the 6 voxels sharing a face with it.

  Relevant Figure Panel(s). Figure 1B.

- $N_{18}$: the intermediate neighborhood in 3D, which is composed of the voxel of interest, the 6 voxels sharing a face with it and the 12 voxels sharing an edge with it.

  Relevant Figure Panel(s). Figure 1B.

- $N_{26}$: the maximum neighborhood in 3D, which is composed of the voxel of interest, the 6 voxels sharing a face with it, the 12 voxels sharing an edge with it and the 8 voxels sharing a vertex with it.

  Relevant Figure Panel(s). Figure 1B.

- Node degree: a integer value reflecting the number of nodes that the node of interest connects to through a single network edge.

  Example(s). In a topological skeleton network, a branch point has a degree of at least three and an endpoint has a degree of one.

- Pixel: the base unit of area in 2D discrete datasets.

- Point: the base unit of a continuous dataset.

- Radius of curvature: the radius of the largest circle that can be fitted to an arc.

  Example. A curved subsegment of an object's perimeter or topological skeleton branch can be described by a radius of curvature.

- Regular polygon: a shape that exists within a single plane of continuous space and is composed of minimum three connected line segments (*i.e.*, sides) of equal length that enclose an area defined by the number and length of line segments.

  Example(s). See definitions for Class 1, Class 2 and Class 3 regular polygons.

The following equations determine regular polygon area ($A$), internal angle ($\theta_i$) and side length ($l$) from their inscribing radius ($r_i$), circumscribing radius ($r_c$) and number of sides ($n$):

$$A = \frac{n}{2} r_i l \quad , \qquad \theta_i = \frac{180(n-2)}{n} \quad , \qquad l = 2r_i \tan\left(\frac{180}{n}\right) = 2r_c \sin\left(\frac{180}{n}\right)$$

Relevant Figure Panel(s). Figure 2A-C and Movie S1-7.

- Roundness ($R$): a dimensionless, 2D shape descriptor used in image processing (*i.e.*, discrete space representation) that compares an object of interest to a perfect circle through a ratio of the object's area ($A$) and perimeter ($P$). Roundness values range between 0 and 1. A roundness value equal to 1 indicates a perfectly circular object. Note: Roundness can alternatively be defined as the ratio between an object's inscribing radius and its circumscribing radius.

$$R = \frac{4\pi A}{P^2}$$

- Set: a collection of unique entities/objects/points/voxels.

  Example(s). See complement definition example.

  Relevant Figure Panel(s). Figure 3B, 5A, 5B, 5D-E', 6A, 6D and 6E.

- Sphericity ($\Psi$): a dimensionless, 3D shape descriptor that compares an object of interest to a perfect sphere through a ratio of the object's surface area ($A$) and it's volume ($V$). Sphericity values range between 0 and 1. A sphericity value equal to 1 indicates a perfectly spherical object.

$$\Psi = \frac{\pi^{1/3}(6V)^{2/3}}{A}$$

- Spine: the longest, or most heavily weighted simple direct path that can be traversed between any two endpoint nodes of a topological skeleton network without using any edges more than once. This is also referred to as the longest, shortest path.

- Thinning algorithm: the process of iteratively removing pixels/voxels from an object's discrete representation until what remains is a 1-pixel/voxel wide connected set of pixels/voxels. Depending on the rules defining the algorithm, the thinned result may or may not reflect the object's original shape and/or the object's original algebraic topology (*e.g.*, Euler characteristic).

  Example(s). See Lam et al., 1992, Sato et al., 2000 and Serino et al., 2011 for descriptions of various thinning algorithms.

- Topological skeleton: the result of algebraic topology preservation and the Medial Axis Transform in discrete space. A set of endpoint, chain and branch point pixels/voxel that reflect all asymmetries in an object's shape. Topological skeletons are generated from binary or labeled instance segmentation digital datasets. Classically, topological skeletons should be single pixel/voxel wide objects; however, this is dependent on the neighborhood kernel and/or thinning algorithm used. It is also possible that algorithms render skeletons with areas of width greater than 1 if the corresponding perimeter/surface portion is highly variable or asymmetric.

  Example. See Lee et al., 1994 for a full description of the topological skeleton generation algorithm used in this paper. See Zhang & Suen, 1984 for the description of another topological skeleton generation algorithm.

  Relevant Figure Panel(s). Figure 1E, 2A, 2C, 2D, 2F, 2G, 3C, 4A', 4C', 4E', 4E'', 4E''', 6D, 6E, 6G-H, Movie S1-11 and Movie S15-18.

- Undirected network graph: a network in which the edges can be traversed in both directions (*e.g.*, you can move from node A to node B over edge 1, and you can move from node B to node A over edge 1).

  Example. In a simple triangular network graph with nodes $A, B, C$, the possible ways to travel over an edge are $A \rightarrow B,\ A \rightarrow C,\ B \rightarrow A,\ B \rightarrow C,\ C \rightarrow A,\ C \rightarrow B$.

  Relevant Figure Panel(s). Figure 1E, 3K, 5A', 5B' and 6A-B.

- Vertex: in discrete space, the base unit of pixel/voxel edge/face bounding as well as a potentially shared entity that can define a neighbor relationship in $N_8$ and $N_{26}$. In continuous space, a point defining the intersection of two or more line segments.

- Voxel: the base unit of volume in 3D discrete datasets.

- Weighted network graph: a network in which the edges and/or nodes are assigned specific weight attributes that make them more or less likely to be part of a path.

  Example. In networks created from topological skeletons, the default network construction encodes a branch's length as its weight. In a network representing connecting flights, and edge can exist between two airports and the number of passengers traveling between two airports can be encoded as the edge's weight.

  Relevant Figure Panel(s). Figure 1E, 3K, 5A', 5B' and 6A-B.